\documentstyle[preprint,aps,epsf]{revtex}

\makeatletter
\def\theequation{\arabic{section}.\theequation@prefix\arabic{equation}}%
\def\mathletters{%
\inc@eqnnum  \setcounter{eqletter}{0}%
\edef\@currentlabel{\theequation}%
\def\theequation{\arabic{section}.\theequation@prefix\arabic{equation}%
           \alph{eqletter}}%
\def\inc@eqnnum{\addtocounter{eqletter}{1}}%
\def\dec@eqnnum{\addtocounter{eqletter}{-1}}%
}%
\def\section{{\setcounter{equation}{0}}
\@mainheadtrue
\@startsection {section}{1}{\z@}{0.8cm plus1ex minus
 .2ex}{0.5cm plus1ex minus.2ex}{\reset@font\small\bf\centering}
           }
\makeatother

\def\svec#1{\skew{-2}\vec#1}

\begin{document}

\baselineskip=14pt plus .5pt minus .5pt


\title{Run-away electrons in \\
relativistic spin ${\bf 1 \over \bf 2}$ quantum
electrodynamics\footnote{This work is supported in part by funds provided by the U.S.
Department of Energy (D.O.E.) under cooperative aggrement \# DF-FC02-94ER40818.}}

\author{F.~E.~Low\\
{\it Center for Theoretical Physics, Department of Physics,}\\
{\it and Laboratory for Nuclear Science,}\\
{\it Massachusetts Institute of Technology,}\\
{\it 77 Massachusetts Avenue,}\\
{\it Cambridge, MA ~02139-4307}\\
\vskip.3in
MIT-CTP-2522 \hfil February 1997}

\maketitle

\thispagestyle{empty}

\bigskip

\begin{abstract}
The existence of run-away solutions in classical and non-relativistic quantum
electrodynamics is reviewed.  It is shown that the less singular high energy behavior of
relativistic spin $1\over2$ quantum electrodynamics precludes an analogous behavior in
that theory.  However, a Landau-like\footnote{L.D. Landau, A.A. Abrikosov, I.M.
Khalatnikov, Doklady Akad.~Nauk.~USSR {\bf 95}, 773 (1954) and {\bf 95}, 1177 (1954).}
anomalous pole in the photon propagation function or in the electron-massive photon
foward scattering amplitude would generate a new run-away, characterized by an
energy scale
$\omega \sim m_e \exp {1 \over \alpha}$.  This contrasts with the energy scale $\omega
\sim {m_e \over \alpha}$ associated with the classical and non-relativistic quantum
run-aways.
\end{abstract}

\bigskip
\section{INTRODUCTION}


\def\threedots#1{\vbox{\ialign{##\crcr%
$\hskip2pt. \hskip-1pt . \hskip-1pt .$\crcr%
\noalign{\kern1pt\nointerlineskip}%
$\hfil\displaystyle{#1}\hfil$\crcr}}}%

Almost a hundred years ago, Lorentz\footnote{H.~A.~Lorentz, {\it Theory of Electrons\/},
2nd edition (Dover, New York, 1952, p.~49); 1st edition, 1909} calculated the self force on
a charged particle, that is, the force exerted on the particle by its own radiation field.  In
an expansion in powers of the radius $b$ of the particle,  he found for the force $\vec{f}$
\begin{equation}
\vec{f} = - \delta m \, \ddot{\vec{y}} + {\textstyle{2\over3}} 
{e^2\over4\pi} \threedots{\vec{y}} + O(b) ~~,
\label{eq:1.1}
\end{equation}
where $\vec{y}$ is the position vector of the particle, 
$\delta m$ is a function of $b$ that goes like $1/b$ for small
$b$, the
$\threedots{\vec{y}}$ coefficient is independent of $b$ 
($e$ is the charge of the particle)
and the higher terms all go to zero as $b \to 0$.  In order to deal with the
problem 
of the divergence of $\delta m$ as $b \to 0$,
Lorentz invented mass renormalization.  Applying Newton's second
law
\begin{equation}
m_0 \ddot{\vec{y}} = - \delta m \ddot{\vec{y}} + 
{\textstyle{2\over3}} {e^2\over4\pi} \threedots{\vec{y}} + O(b)
\label{eq:1.2}
\end{equation}
or, with $m = m_0 + \delta m$ fixed and $b \to 0$,
\begin{equation}
m \ddot{\vec{y}} = {\textstyle{2\over3}} {e^2\over4\pi} \threedots{\vec{y}} \ \ ,
\label{eq:1.3}
\end{equation}
a finite equation of motion.  However, the divergence problem returns
through the back door: (\ref{eq:1.3}) has a run-away solution
\begin{equation}
\vec{y} = \vec{y}_0 + \left( \vec{v}_0 - \vec{a}_0  {e^2 \over 6\pi m}  \right) t + 
\vec{a}_0 \left( {e^2 \over 6\pi m} \right)^2
\,  \left( {\rm exp}{6\pi m t \over e^2} -1 \right)
\label{eq:1.4}
\end{equation}
with a run-away time of $\tau = {2\over3} \cdot {e^2 \over 4\pi mc^3}$,
$\sim 10^{-23}$ seconds for electrons.

A Lorentz covariant equation which reduces to (\ref{eq:1.3}) in the instantaneous rest
system of the particle was proposed by Dirac \footnote{P.~A.~M.~Dirac, {\it
Proc.~Roy.~Soc.\/}~{\bf A167}, 148 (1938)}; however, the run-away problem was
not removed.  Of course, in the relativistic model it is the momentum, not the
velocity, that runs away.

The clearest algebraic insight into the problem, both classical and
quantum, is obtained by finding an upper half-plane pole in the Laplace transform 
$\vec{y}(\omega)$ of $\vec{y}(t)$:
\begin{equation}
\vec{y}(\omega) = \int_0^\infty dt \, e^{i \omega t} \, \vec{y}(t) ~~.
\label{eq:1.5}
\end{equation}
Here $\omega = \omega_1 + i \omega_2$, with $\omega_2$ positive and
large enough to make the integral converge.
In Appendix A we take that path and, following Moniz and Sharp 
\footnote{E.~J.~Moniz and D.~H.~Sharp, {\it Phys.~Rev.\/}~{\bf D10}, 1133 (1974);
                            {\it Phys.~Rev.\/}~{\bf D15}, 2850 (1977)},~
show that in a suitably cut-off classical theory
there are no run-aways, no matter how small the cut-off
radius, provided $m_0$, the bare mass, is positive 
We then show that the mass renormalization,
carried out for $b \to 0$ as described above,
leads to run-aways.

In Appendix B, we obtain the same result in a non-relativistic quantum theory.  Thus
quantum effects per se do not cure the disease, nor does the addition of a confining
oscillator potential.

In Appendices A and B we considered, and in the following we consider,
the simplest problem that shows the 
run-away
effect:  a charged (in the limit)
point particle, which we shall call an electron, is in a wave-packet
state with approximate momentum $\vec{p}$ and energy $E$ for times $t
\leq 0$.  At $t=0$, a weak spatially uniform external electric field $\vec{E}^e
(t)$ is switched on, and we follow the electron's motion, calculating the
mean value of the coordinate $\vec{y}$, or of the velocity
$\vec{v}(t) = \dot{\vec{y}}(t)$.

The solution for $\vec{y}(\omega)$ is given for classical motion in (\ref{eq:A.19}) of
Appendix A and for the non-relativistic quantum case by (\ref{eq:B.16}) of Appendix B:
\begin{equation}
\vec{y} (\omega) = -
{ e \vec{E}^e (\omega) / \omega^2 \over m_0 + {2\over3} e^2 
{1\over (2\pi)^3}
\int {d \vec{k} \over k^2 - \omega^2} \, 
| f(\vec{k}) |^2} ~~.
\label{eq:1.6}
\end{equation}
Here $f(\vec{k})$ is the cut-off function, which must go to zero as
$k\to\infty$ fast enough to make the integral converge.
$\vec{E}^e(\omega)$ is the Laplace transform of $\vec{E}^e(t)$,
$$
\vec{E}^e(\omega) = \int_0^\infty dt \, e^{i\omega t} \vec{E}^e(t) ~~,
$$
and $\vec{y} (\omega)$ in the quantum case represents the mean value of the variable.
The absence (in (\ref{eq:1.6}) of an upper half $\omega$ plane pole in the cut-off theory,
and its emergence following Lorentz's renormalization procedure, are shown
in Appendix A.  The presence of such a pole in the Laplace transform
produces an exponential run-away in time, via the inversion formula
\setcounter{equation}{6}
\begin{equation}
\vec{y}(t) = {1\over 2\pi}
\int_{-\infty+i \omega_2}^{+\infty+i \omega_2}
d \omega \, e^{-i \omega t} \vec{y} (\omega) ~~.
\label{eq:1.7}
\end{equation}

The upper half plane pole appears at a frequency $\hbar \omega = {3i \over 2} \, {mc^2
\over \alpha}$, where $i = \sqrt{-1}$ and $\alpha$ is the fine structure constant. 
We point out here that  for weak coupling the energy $m \over \alpha$ is deep in
the relativistic region where the equations used to describe the system are certainly
not valid.  Since relativistic spin ${1 \over 2}$ quantum electrodynamics is known to hold
accurately in that energy region, we presume that there will be no analogous pole in that
theory.  In the following we formulate a way of studying that issue, and indeed find no
analogous pole in the relativistic spin ${1 \over 2}$ theory.  However, if a pole of the
type suggested by Landau, Abrikosov and Khalatnikov exists, either in the photon
Green's function or in the electron photon forward scattering amplitude, there would be a
new type of run-away with the characteristic frequency
$\hbar \omega \sim imc^2 {\rm exp} {1/\alpha}$, high enough so that there appears to
be no way in which present day experiments can rule it out.  Of course, such a pole is
already a symptom of a diseased theory.

In the following sections we will be calculating the mean value of the electron velocity,
to lowest order in the external field.  Of course, the limitation of the electron's velocity to
$c$ does not automatically occur in this approximation.  However, if the electron
momentum does run away, the limitation of the velocity to $c$ will appear as a
non-linear effect of the external field.  Correspondingly, the linear approximation to a
run-away momentum would be a run-away velocity, for which we search.

\goodbreak\bigskip

\section{SPINOR QUANTUM ELECTRODYNAMICS}

We work in the Schroedinger representation with a state vector $\Psi(t)$
which satisfies the equation 
\begin{equation}
-{1 \over i} {\partial \Psi \over \partial t}
= (H + H') \Psi ~~.
\label{eq:2.1}
\end{equation}
Here $H$ is the electron-photon Hamiltonian, and $H'$ the interaction
Hamiltonian with the spatially uniform external field:
\begin{equation}
H' = - \int \vec{\jmath} (\vec{x}) \cdot \vec{A}^e (t) d\vec{x}
\label{eq:2.2}
\end{equation}
where
\begin{equation}
\vec{E}^e = - {\partial \vec{A}^e \over \partial t}
\label{eq:2.3}
\end{equation}
and 
$\vec{\jmath} (\vec{x})$ is the electron current density
\begin{equation}
\vec{\jmath} (\vec{x}) 
= e \psi^\dagger (\vec{x}) \vec{\alpha} \psi(\vec{x})
\label{eq:2.4}
\end{equation}
where $e,~\psi,~\vec{\alpha}~{\rm and}~\vec{x}$ have their usual
meaning.

We take for the velocity operator 
\begin{equation}
\vec{v} = {\int d\vec{x}' \, \vec{\jmath} (\vec{x}') \over
           \langle \, \int d\vec{x} \, \rho         (\vec{x}) \, \rangle}
\label{eq:2.5}
\end{equation}
where $\rho$ is the charge density%
\footnote{Charge renormalization must be taken into account here.
This is discussed in Section III, Eq.~(\ref{eq:3.16}).}
\begin{equation}
\rho = e \psi^\dagger (\vec{x}) \psi(\vec{x}) ~~.
\label{eq:2.6}
\end{equation}
For $t < 0$, the electron is in a state $\Psi_0$ which satisfies the
equation 
\begin{equation}
- {1\over i} \, {\partial \Psi_0 \over \partial t} = H \Psi_0 ~~.
\label{eq:2.7}
\end{equation}
$\Psi_0$ may be expressed as a superposition of stationary states of
momentum $\vec{p}$ and spin $s$:
\begin{equation}
\Psi_0 = \sum_{s} \int \phi(\vec{p},s) \, \Psi_{\vec{p},s} \, d\vec{p}
\label{eq:2.8}
\end{equation}
where
\begin{equation}
H \Psi_{\vec{p},s} = E(\vec{p}) \Psi_{\vec{p},s} ~~. \footnotemark[6]
\label{eq:2.9} 
\end{equation}
\footnotetext[6]{We ignore problems associated with infra-red singularities.}
The state vector normalization is
\begin{equation}
(\Psi_{p',s'}, \Psi_{p,s}) 
= \delta s' s (2\pi)^3 \delta (\svec{p}' - \vec{p}) ~~,
\label{eq:2.10}
\end{equation}
so that the wave-packet state $\Psi_0$ is normalized to one with 
\begin{equation}
\sum_{s} \int d\vec{p} | \phi (\vec{p},s) |^2 \cdot (2\pi)^3 = 1 ~~.
\label{eq:2.11}
\end{equation}
The mean value of the velocity vector in the state (\ref{eq:2.8}) is
\begin{equation}
\sum_{s',s} \int \phi^{*} (\svec{p}',s') d \svec{p}'
\left(
\Psi_{\svec{p}',s'} ~,~ 
{\vec{\jmath}(\vec{x}) \over \int d\vec{x}' \langle \rho \rangle} 
\, \Psi_{\vec{p},s}
\right)
\times
\phi(\vec{p},s) d\vec{p} \, d \vec{x} ~~.
\label{eq:2.12}
\end{equation}
The matrix element of $\int \vec{\jmath} (\vec{x}) d\vec{x}$ is
\begin{equation}
\int \left(
\Psi_{p',s'} ~,~ {\vec{\jmath} (\vec{x})}
\, \Psi_{p,s} \right) d\vec{x}
= (2\pi)^3 \delta (\svec{p}' - \vec{p}) u^{*} (\svec{p}',s') \,
e \vec{\alpha} \, u(\vec{p},s) ~~;
\label{eq:2.13}
\end{equation}
since $\svec{p}' = \vec{p}$, there is no form factor or anomalous
magnetic term.
The spinors $u(\vec{p},s)$ satisfy the Dirac equation
\begin{equation}
\left( \vec{\alpha} \cdot \vec{p} + \beta m \right)
\, u(\vec{p},s) = E(\vec{p}) \, u(\vec{p},s) ~~.
\label{eq:2.14}
\end{equation}
The matrix-element in (\ref{eq:2.13}) has the value
\begin{equation}
u^* (\vec{p},s') \vec{\alpha} \, u(\vec{p},s) = \delta_{s's} ~\vec{p}/E 
~~,
\label{eq:2.15}
\end{equation}
giving for the mean value (\ref{eq:2.12})
\begin{equation}
(\Psi_0, \vec{v} \Psi_0) 
= \sum_s \int | \phi (\vec{p},s) |^2 \, \vec{p}/E 
{~} d\vec{p} \, (2\pi)^3 ~~.
\label{eq:2.16}
\end{equation}
We can, in this calculation and others that involve momentum conserving
operators, dispense with the wave packet $\phi$, replace the state
$\Psi_0$ by $\Psi_{\vec{p}}$, and leave the $d\vec{x}$ integral undone,
thereby eliminating
the delta function in (\ref{eq:2.13}), and replacing the wave packet
average in (\ref{eq:2.16}) by the single value $\vec{p}/E$.  From here on we will assume
that this has been done.  For the real electron, the wave-packet average must be
taken.

We solve (\ref{eq:2.1}) for small $\vec{A}^e$ by expanding in powers of
$\vec{A}^e$.  With 
\begin{equation}
\Psi = \Psi_0 + \Psi_1 + \ldots
\label{eq:2.17}
\end{equation}
we have
\begin{equation}
-{1\over i} \, {\partial \Psi_0 \over \partial t} = H \Psi_0 
\label{eq:2.18}
\end{equation}
and
\begin{equation}
-{1\over i} \, {\partial \Psi_1 \over \partial t} = H \Psi_1 + H' \Psi_0
\label{eq:2.19}
\end{equation}
so that
\begin{equation}
\Psi_1 = -i \int_0^t dt' \, e^{i H (t'-t)} H'(t') \Psi_0 (t') ~~.
\label{eq:2.20}
\end{equation}
As just described, we replace $\Psi_0$ by $\Psi_{\vec{p},s}$
and leave the $d\vec{x}$ integral in (\ref{eq:2.5}) undone.  The mean value of
the velocity is then given by 
\begin{equation}
\langle \vec{v} \rangle = \vec{p}/E + \delta \vec{v} 
\label{eq:2.21}
\end{equation}
with
\begin{equation}
\delta\vec{v} = -i \left( \Psi_p ~,~ \vec{v} \int_0^t d t' 
\, e^{i H(t'-t)} H'(t') \Psi_p \right) + {\rm ~c.c.}
\label{eq:2.22}
\end{equation}
for 
$t > 0$ and $\delta \vec{v}$ understood to be zero for $t < 0$.

We make (\ref{eq:2.22}) more explicit:
\begin{equation}
\delta v^\ell = i \left( \Phi_p ~,~ v^\ell \int_0^t dt' \,
e^{i (H-E) (t'-t)} j^k (\svec{x}') \Phi_p \right)  A^e_k(t') d\svec{x}'
+ {\rm ~c.c.}
\label{eq:2.23}
\end{equation}
where $\Phi_p = \Psi_p (t=0) = e^{i E_p t} \Psi_p$.

To compare with the results from Appendices A and B we take the Laplace
transform of (\ref{eq:2.23}).
We note first that the Laplace transform of
$\int^t_0 dt' e^{i (H-E)(t'-t)} A^e_k(t')$ is
\begin{eqnarray}
\int_0^\infty dt & \, e^{i\omega t}&\int_0^t dt' \, 
e^{i(H-E)(t'-t)} A^e_k(t') \nonumber \\ 
&=& \int_0^\infty dt' \, e^{i\omega t'} A^e_k(t')
\, \int_{t'}^\infty dt \, e^{i\omega(t-t')} e^{i(H-E)(t'-t)} \label{eq:2.24}\\
&=& -{1 \over i(\omega-(H-E))} A^e_k (\omega) \nonumber
\end{eqnarray}
(where $A^e_k(\omega)$ is the Laplace transform of $A^e_k(t)$)
so that the Laplace transform in (\ref{eq:2.23}) is
\begin{eqnarray}
\delta v^\ell (\omega) &=&
- \left( \Phi_p ~,~ v^\ell {1\over \omega-(H-E)}   j^k ({\vec{x}}') \Phi_p
\right) A^e_k(\omega)
{~+~} {\scriptstyle{\rm contribution~of~complex~conjugate}}
\label{eq:2.25}\\
\noalign{\hbox{or~in~all}}
\delta v^\ell (\omega) &=& -\left( \Phi_p ~,~ 
\left[ v^\ell {1\over \omega-(H-E)}   j^k ({\vec{x}}') -   j^k
({\vec{x}}') {1\over
\omega+H-E} v^\ell
\right] \Phi_p \right) A^e_k(\omega) ~~.
\label{eq:2.26}
\end{eqnarray}
If we took the expression (\ref{eq:2.26}) at face value, we would immediately conclude
that there could be no upper half plane singularities in $\delta v^\ell (\omega)$, since $H$
has only real eigenvalues.  However, since the theory is not finite, we are not justified in
drawing that conclusion.  We must rather deal with the mass and charge renormalized
theory, which in practise means order by order perturbation theory in $\alpha$.  In the
next section we will show how the expression $\delta v^\ell (\omega)$ in (\ref{eq:2.26})
can be calculated order by order from Feynman diagrams.

Before we turn to that, we exploit (\ref{eq:2.26}) to calculate the zero'th order
(in the fine structure constant) motion of the electron in the weak external
field.\footnotemark[7]
\footnotetext[7]{We will see in Appendix C and Section III how the calculation can
be done directly from Feynman diagrams.}  There are two classes of intermediate states: 
one particle (which makes no contribution), and three particles (two electrons and a
positron).  One finds for
$\delta v^\ell (\omega)$
\begin{equation}
\delta v^\ell (\omega) = -e
\left[
{u^* \alpha^\ell \Lambda^{+} (\vec{p}) \alpha^k u \over \omega}
-
{u^* \alpha^k \Lambda^{-} \alpha^\ell u \over \omega - 2E}
+ \omega \leftrightarrow -\omega, \ell \leftrightarrow k 
\right]
\, A^e_k(\omega) ~~.
\label{eq:2.27}
\end{equation}
Here $\Lambda^{+}(\svec{p})$ is the positive energy projection operator
with momentum $\svec{p}$, 
$\Lambda^-(\svec{p})$ the negative energy projection operator of momentum
$\svec{p}$ (corresponding to positron momentum $-\svec{p}$):
\begin{equation}
\Lambda^\pm (\svec{p}) 
= {E \pm (\vec{\alpha} \cdot \vec{p} + \beta m) \over 2E} ~~.
\label{eq:2.28}
\end{equation}
The first term in (\ref{eq:2.27}) is symmetric in $\ell$ and $k$, and hence gives
zero when $\omega \leftrightarrow -\omega$.  To calculate the second
term, we need 
\begin{eqnarray}
u^* \alpha^k \Lambda^{-} \alpha^\ell u &=&
u^* \alpha^k 
\left( {\textstyle{1\over2}} - {H_D \over 2E} \right)
\alpha^\ell u \nonumber \\
&=& u^* \left( \alpha^k \alpha^\ell - {p^k \over E} \alpha^\ell \right) u \nonumber \\
&=& \left( \delta^{kl} - {p^k p^\ell \over E^2} \right)
+ i \sigma_{kl} 
\label{eq:2.29}\\
{\rm where~~~~~~~~} i \, \sigma_{kl} &= 
{\alpha^k \alpha^\ell - \alpha^\ell \alpha^k \over 2} 
\label{eq:2.30}
\end{eqnarray}
so that
\begin{equation}
\delta v^\ell (\omega) = e \left\{
+ \left( \delta^{kl} - {p^k p^\ell \over E^2} \right)
\left( {1\over \omega-2E} - {1\over \omega+2E} \right)
+ i \, <\sigma_{kl} >
\left( {1\over \omega-2E} + {1\over \omega+2E} \right)
\right\} A^e_k(\omega) ~~.
\label{eq:2.31}
\end{equation}

We see in (\ref{eq:2.31}) the structure that will guide us in the search for run-away
poles.  The amplitudes that are odd in interchange of $k$ and $\ell$ are odd in
$\omega$, to all orders in $\alpha$; those that are even in $(k,\ell)$ interchange are
even in $\omega$ to all orders in $\alpha$.  Higher order odd amplitudes in spinor
electrodynamics go like ${1 \over \omega}$ (at high $\omega$) except for
logarithms; higher order even amplitudes go like ${1 \over \omega^2}$ except for
logarithms.  There is therefore no way for a second order term to be of the same order
as a zero'th order term (the sign of a possible pole) at $\omega \sim {m \over \alpha}$, as
in the non-relativistic case; rather, it must be of order $\omega \sim m{\rm exp}(1/
\alpha)$.  Indeed, the pole$^1$ conjectured by Landau, Abrikosov and Khalatnikov would
arise in just that way.

We translate (\ref{eq:2.31}) into time dependence:
\begin{equation}
\delta v^\ell (t) = {1\over 2\pi}
\int\limits_{-\infty+i\omega_2}^{\infty+i\omega_2}
{~} d\omega {~} e^{-i \omega t} \delta v^\ell (\omega) ~~,~~~~ t \geq 0 ~~.
\label{eq:2.32}
\end{equation}
We close the $\omega$ integration in the upper or lower half plane as
called for, and find
\begin{eqnarray}
\delta v^\ell (t) &=& \left( \delta^{kl} - {p^k p^\ell \over E^2} \right)
{e \over E} \int_0^t \left\{ 1 - \cos [2E(t'-t)] \right\} E^e_k (t') dt' \nonumber \\
&& \hbox{\qquad} + e \, \sigma_{kl} \, {1\over E} \int_0^t \sin [2E(t'-t)]
E^e_k (t') dt' ~~.
\label{eq:2.33}
\end{eqnarray}

We recognize in (\ref{eq:2.33}) the classical relativistic first order equation for $\delta
v^\ell$.  This follows by neglecting the  zitterbewegung like sine and
cosine averages of the external electric field --- a valid neglect for
fields whose characteristic time is much longer than $\hbar / E$ (which
is $\sim 10^{-21}$ seconds for electrons).

Finally, we wish here to find the $\omega$ dependence of $\delta v^\ell$
at high $\omega$.  From (\ref{eq:2.26}), it is
\begin{eqnarray}
\delta v^\ell(\omega) &\to&
- {1\over\omega} \left( \Phi_p ~,~ [v^\ell, j^k] \Phi_p \right) \, A^e_k(\omega)
\label{eq:2.34}\\
&=& - {2i e A^e_k(\omega)\over\omega} 
\left( \Phi_p ~,~ \psi^{\dagger}(x) \, 
\sigma^{\ell k} \psi(x) \Phi_p \right) ~~.
\label{eq:2.35}
\end{eqnarray}

The matrix element in (\ref{eq:2.35}) is of course finite
in zero'th order, as shown in (\ref{eq:2.31}).  In second order, it would be
finite except for vacuum polarization effects.  We shall see in Section
III that the correctly renormalized Feynman amplitude for $\delta
v_\ell$ implies that the expression (\ref{eq:2.26}) must carry an explicit factor
${1\over Z_3^{\,2}}$, where $Z_3$ is the photon propagator
renormalization constant
\begin{equation}
Z_3 = 1 - {\alpha \over 3\pi} \log {\Lambda^2 \over m^2} +
{\scriptstyle{\rm finite}} + 
{\scriptstyle{\rm higher~order~in~}} \alpha
\label{eq:2.36}
\end{equation}
with $\Lambda$ a momentum cut-off.  The finiteness of (\ref{eq:2.26}) together
with the divergence in
$Z_3^{\,-2}$ implies a high
$\omega$ dependence,
\begin{equation}
\delta v^\ell \to - {2\over\omega} \, i e A^e_k(\omega)
\psi^\dagger \sigma^{\ell k} \psi
\left( 1 + 2 {\alpha \over 3\pi} \log {\omega^2 \over m^2} \right) ~~.
\label{eq:2.37}
\end{equation}

Using the technique discussed in Section III, we can calculate the high
$\omega$ behavior of $\delta v^\ell$ explicitly, and find agreement with
 (\ref{eq:2.37}). 

\goodbreak\bigskip

\section{LOWEST ORDER RADIATIVE CORRECTION}

We consider first the mean value of the velocity operator, without the
accelerating electric field.  The stationary state $\Phi_p$ is given as
\begin{equation}
\Phi_p = U(0,-\infty) \chi_p
\label{eq:3.1}
\end{equation}
where $\chi_p$ is the bare (or interaction representation) one particle
state, and $U(t_1,t_2)$ is the unitary operator which transforms that
state in time:
\begin{eqnarray}
- {1\over i} \, {\partial \over \partial t_2} \, 
U(t_2,t_1) &=& H_I(t_2) \, U(t_2,t_1)
\label{eq:3.2}\\
\noalign{\noindent{\rm and}}
U(t_1,t_1) &=& 1 ~~.
\label{eq:3.3}
\end{eqnarray}
$H_I$ is the interaction Hamiltonian expressed in terms of interaction
representation operators:
\begin{equation}
H_I = \int d \vec{x} \left[ - e A_\mu (x) \overline{\psi} (x)  i \gamma^\mu \psi(x) -
\delta m \overline{\psi}(x) \psi(x) \right]~~.
\label{eq:3.4}
\end{equation}
The free Hamiltonian $H_0$ is expressed in terms of the correct mass
of the electron; hence the necessity for the subtraction of
$\delta m \overline{\psi} \psi$ in (\ref{eq:3.4}).

We wish to calculate the velocity operator in the Schroedinger state
$\Psi_p$.  It is
\begin{eqnarray}
\langle v^k \rangle &=& (\Psi_p, v^k \Psi_p) \nonumber \\
                    &=& (\Phi_p, e^{iHt} v^k e^{-iHt} \Phi_p) \nonumber \\
&=& \left( U(0,-\infty) \chi_p ,~ e^{iHt} e^{-iH_0 t} v^k(t) e^{i H_0 t} 
e^{-iHt} U(0,-\infty) \chi_p \right) ~. 
\label{eq:3.5}
\end{eqnarray}
Here we recognize the operator
\begin{equation}
e^{i H_0 t} e^{-i H t} = U(t,0)
\label{eq:3.6}
\end{equation}
so that $v^k(t)$ is the velocity expressed in the interaction
representation, and 
\begin{equation}
\langle v^k \rangle = 
\left( S \chi_p , U(\infty,t) v^k(t) U(t,-\infty) \chi_p \right)
\label{eq:3.7}
\end{equation}
where the $S$ matrix $S = U(\infty,-\infty)$, and where we have used the
unitarity of $U$ and its group property:
\begin{equation}
U(t_3,t_2) \, U(t_2,t_1) = U(t_3,t_1) ~~.
\label{eq:3.8}
\end{equation}
The $S$ matrix is diagonal on one particle states.  Since there is no
self-energy correction, it must be a phase, which is canceled by
disconnected diagrams. 


In (\ref{eq:3.7}) we see $\langle v^k \rangle$ expressed in a way which allows us
to make use of the Dyson technique and Feynman diagrams.  
The 2nd order result is expressed in the diagrams shown in Fig.~1.

\def\lfiga{\vcenter{\hbox to 24pt{\hss\vbox to 60pt{\vss
\vss}\hss}}}

\def\lfigb{\vcenter{\hbox to 24pt{\hss\vbox to 60pt{\vss
\vss}\hss}}}

\def\lfigc{\vcenter{\hbox to 24pt{\hss\vbox to 60pt{\vss
\vss}\hss}}}

\def\lfigd{\vcenter{\hbox to 24pt{\hss\vbox to 60pt{\vss
\vss}\hss}}}

$$
\lfiga ~~+~~ \lfigb ~~+~~ \lfigc ~~+~~ \lfigd
$$
\centerline{
$\hbox to 24pt{\hss\hbox{(a)}\hss} ~~~~~~~
\hbox to 24pt{\hss\hbox{(b)}\hss} ~~~~~~~
\hbox to 24pt{\hss\hbox{(c)}\hss} ~~~~~~~
\hbox to 24pt{\hss\hbox{(d)}\hss}$}
\centerline{Fig.~1}

Here the initial and final lines represent the wave functions
$\overline{u}(p) \ldots u(p)$; the dot $\bullet$ 
represents the operator $i \gamma^\mu$;
and so forth.  The diagrams (1b) and (1c) have the
$\delta m$ correction subtracted from them, and are multiplied by 
${1\over2}$ to reproduce the correct normalization of the incoming
state.  The sum of all such diagrams is
\begin{equation}
\langle v^\mu \rangle = \sqrt{Z_2} \, \overline{u} \, {1\over Z_1}
i \gamma^\mu u \, \sqrt{Z_2} = \overline{u} \, i \gamma^\mu u
\label{eq:3.9}
\end{equation}
since $Z_2 = Z_1$.  Thus, $\langle v^\mu \rangle$ is both ultra-violet
and infra-red finite, and equal to its expected free particle value,
\begin{equation}
\langle v^\mu \rangle = {p^\mu \over E}
= \left( {\vec{p} \over E}, 1 \right) ~~.
\label{eq:3.10}
\end{equation}
The closed fermion loop diagrams cancel because $\vec{\jmath}$ and
$\rho$ carry the same charge renormalization.

We turn next to the effect of the external field as given by (\ref{eq:2.23}).  

We consider the Feynman amplitude
\begin{equation}
G_{\,+}^{\ell k} (q^0) = 
i \int^\infty_{-\infty} dx^0 e^{iq^0 x^0} 
\int d\vec{x}
\left( \Phi_p, \left( j^\ell  (\vec{x}, x^0) j^k (\vec{y}, 0)\right)_+  \Phi_p \right)
\label{eq:3.11}
\end{equation}
which we can readily calculate, order by order, following Dyson:
\begin{equation}
G_{\,+}^{\ell k} (q^0) = 
i \int^\infty_{-\infty} dx^0 e^{iq^0 x^0} 
\int d\vec{x}
\left( \chi_p, \left( U (\infty,-\infty)  j^\ell  (\vec{x}, x_0) j^k (\vec{y}, 0)\right)_+  \chi_p
\right) ~~.
\label{eq:3.12}
\end{equation}
$G_{\,+}^{\ell k} (q^0)$ in (\ref{eq:3.11}) and (\ref{eq:3.12}) is the forward scattering
amplitude of a photon of four momentum $q^\mu = (q^0, \vec{0})$ on an electron of
momentum $\vec{p}$.  By considering the form (\ref{eq:3.11}) we see that 
\begin{equation}
G_{\,+}^{\ell k} (q^0) = 
- \left( \Phi_p, \int d\vec{x} j^\ell (\vec{x},0) {1 \over q^0 + i\epsilon - (H-E)} 
j^k (\vec{y},0) -  j^k (\vec{y},0) {1 \over q^0 -i\epsilon + (H-E_0)} 
\int  d\vec{x} j^\ell (\vec{x},0) \Phi_p
\right)
\label{eq:3.13}
\end{equation}
which has the same structure as (\ref{eq:2.26}) except that $q^0 -i\epsilon$ in the
second term of (\ref{eq:3.13}) is in the lower half plane, whereas $\omega$ in
(\ref{eq:2.26}) is in the upper half plane.  We therefore obtain the correct formula
(\ref{eq:2.26}) from (\ref{eq:3.12}) by continuing $q^0$ into the upper half plane and
setting it equal to $\omega$.

The use of Feynman diagrams makes clear the finiteness and high $q^0$
behavior of (\ref{eq:2.26}).  We illustrate with the 2nd order (in $\alpha$)
terms.  The diagrams (with the exception of vacuum polarization) are:


\def\tfiga{\vcenter{\hbox to 30pt{\hss\vbox to 60pt{\vss
\vss}\hss}}}

\def\tfigb{\vcenter{\hbox to 30pt{\hss\vbox to 60pt{\vss
\vss}\hss}}}

\def\tfigc{\vcenter{\hbox to 30pt{\hss\vbox to 60pt{\vss
\vss}\hss}}}

\def\tfigd{\vcenter{\hbox to 30pt{\hss\vbox to 60pt{\vss
\vss}\hss}}}

\def\tfige{\vcenter{\hbox to 30pt{\hss\vbox to 60pt{\vss
\vss}\hss}}}

\def\tfigf{\vcenter{\hbox to 30pt{\hss\vbox to 60pt{\vss
\vss}\hss}}}

$$
\tfiga {~~+~} \tfigb {~+~} \tfigc {~+~}
\tfigd {~+~} \tfige {~+~~} \tfigf {~+~}
{\rm crossed \atop diagrams} ~.
$$
\vskip-8pt\centerline{$%
\hskip6pt\hbox to 30pt{\hss\hbox{(a)}\hss} \phantom{~~+~}
\hbox to 30pt{\hss\hbox{(b)}\hss} \phantom{~+~}
\hbox to 30pt{\hss\hbox{(c)}\hss} \phantom{~+~}
\hbox to 30pt{\hss\hbox{(d)}\hss} \phantom{~+~}
\hbox to 30pt{\hss\hbox{(e)}\hss} \phantom{~+~~}
\hbox to 30pt{\hss\hbox{(f)}\hss} \phantom{~+~}
\phantom{{\rm crossed \atop diagrams} ~.}~~~~~~$}
\centerline{Fig.~2}

\message{hey!}

In Figs.~(2a) and (2c), the self-mass is subtracted, and the remainder
diagrams multiplied by ${1\over2}$.  In Fig.~(2b), the self-mass is
subtracted.  In what remains, the ultra-violet divergences cancel
(using $Z_2=Z_1$ again)
by adding
$$
{\textstyle{1\over2}} (2a) + 
{\textstyle{1\over2}} (2c) + 
(2b) + (2d) + (2e).
$$
The infra-red divergence cancels by adding
${1\over2} (2a) + {1\over2} (2b) + (2f)$.  The high $q^0$ dependence of each of the
diagrams (1b), (1d), (1e) and (1f) is ${1 \over q^0} \log q^0$.  However, when the four
diagrams are added the logarithms cancel, leading to a high $\omega$ dependence in the
odd amplitudes
$$
v^\ell (\omega) \sim {1 \over \omega}
$$
which therefore cannot in weak coupling produce the upper half plane singularity which
classically leads to the run-away solutions.  Note however that the above simple
asymptotic calculation only shows a cancellation in the $(k,\ell)$ odd amplitude.  The
logarithms in the $(k,\ell)$ even amplitudes fail to cancel, leaving the possibility of a
singularity of order $\omega \sim me^{1 \over \alpha}$ in the even amplitudes, and a
run-away $x \sim {\rm exp}(me^{1 \over \alpha} t)$ in  the even amplitudes.  We should
call this a Landau run-away, as opposed to a Lorentz run-away.  The second order results
of these calculations are given in Appendix C.

The relativistic theory brings in a new effect, the electron-positron vacuum
polarization.  To take this into account we must add two more diagrams to those of Fig.~2:
\centerline{\epsffile{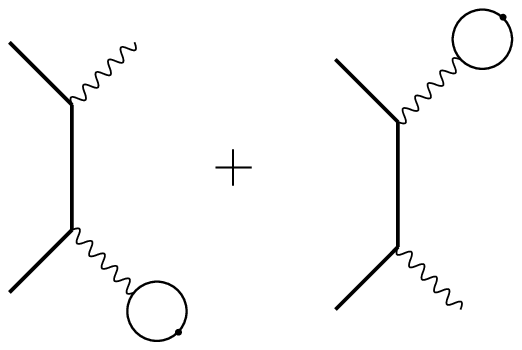}}
\centerline{Fig.~3}

We study the effect of the diagrams of Fig.~3 on the renormalization of (\ref{eq:2.26}) 
We replace $v^\ell $ by
${e_0 \int d \vec{x} L^\ell (\vec{x},0)\over < \rho >}$
where $e_0$ is
the bare charge and $L^\ell(\vec{x},t)$ the lepton current:
\begin{equation}
L^\ell(\vec{x},t) = \psi^{+} (\vec{x}) \alpha^\ell \psi(x) ~~.
\label{eq:3.14}
\end{equation}
Eq.~(\ref{eq:2.26}) becomes
\begin{equation}
\delta v^{\ell} (\omega) = -\left( \Phi_p, \left[
\int {L^\ell (\vec{x},0) d\vec{x} \over <\rho>/e_o} \,\,{1 \over \omega - (H-E)} 
{e_0 L^k (\vec{y},0) \atop {\phantom 1}} + {\rm
crossed~term}
\right]
\Phi_p \right) A_k^{e,u}
\label{eq:3.15}
\end{equation}
where $A_k^{e,u}$ is the unrenormalized external vector potential.

In the combination ${L^\ell \over <\rho>/e_o}$ the numerator operator will produce a
factor
$Z_3 q^2 D_{FC} (q^2)$, where $D_{FC}$ is the finite photon propagation function,
normalized to ${1 \over q^2}$ as $q^2 \to 0$.  The denominator is $Z_3$.  The
calculation of the numerator matrix element will automatically generate the
factor $Z_3$; the denominator must be put in by hand.

On the right side, the unrenormalized external field is produced by an external current
with a charge $e_0$.  The renormalized external field would be produced by a charge
$e$.  Therefore $A_k^{e,u} = {e_0 \over e} A_k^e$, where $A_k^e$ is the renormalized
external potential.  The matrix element containing $e_0 L^k A_k^{e,u} = e_0^2 {L^k \over
e} A_k^e$ will therefore produce a finite factor $q^2 {D_{FC} (q^2) A^e_k \over e} e^2$;
correspondingly the matrix element of $L^k$ will produce a renormalization factor $q^2
D_{FC} (q^2) Z_3$.  We may therefore rewrite (\ref{eq:3.15}) as the finite
expression
\begin{equation}
\delta v^{\ell} (\omega) = - {e \over Z_3^2} \left( \Phi_p, \left[
L^\ell  {1 \over \omega - (H-E)} L^k + {\rm crossed~term} \right]
\Phi_p \right) A_k^e \ \ ,
\label{eq:3.16}
\end{equation}
justifying the discussion following Eq.~(\ref{eq:2.35}).

It is straightforward to show directly from the diagrams in Fig.~3 that the logarithmic
dependence of $D_{FC} (q^2)$ at high $q^2$ translates into the expected logarithmic
behavior of $\delta v^\ell (\omega)$, as given in (\ref{eq:2.36}) and (\ref{eq:2.37}).  (See
Appendix C.)

It is possible that the series of radiative corrections to $D_F$ sum to a pole, as
conjectured by Landau, Abrikosov and Khalatnikov$^1$.  If so, the pole is at space like
$q^2$, {\it i.e.} $(q^{0})^2 <0$ and $\omega^2 <0$, producing a new class of run-away
solutions.

We emphasize that the above discussion is specifically for spin
${1\over2}$ particles.  What happens to the classical run-away solutions
in the relativistic quantum electrodynamics of a spin zero particle is
an interesting question, but has not been dealt with here.

\bigskip

\centerline{\bf Acknowledgment}

I wish to thank Arthur Kerman for arousing my interest in this subject,
and to thank him, Kenneth Johnson and Al Mueller for very helpful
discussions.

\newpage

\def\theequation{A.\arabic{equation}}%
\appendix
\section{Classical run-away}

We consider the simplest problem:
an electron is placed at $\vec{y} = 0$ with $\dot{\vec{y}}=0$.  
At $t=0$, an external spatially constant electric field $\vec{E}^e(t)$
is turned on.  We solve the equations of motion exactly in the weak
$\vec{E}^e$ limit.  We work with a cut-off theory, with a cut-off
function $f(\vec{x})$.  The equations of motion are
\begin{eqnarray}
m_0 \ddot{\vec{y}} &=& \int e
\left[ \vec{E} (\vec{x},t) + \dot{\vec{y}} \times \vec{B}(\vec{x},t) 
+ e \vec{E}^e(t)
\right] 
f(\vec{x}-\vec{y}) \, d\vec{x} ~~,
\label{eq:A.1} \\
\nabla\times\vec{E} &=& - {\partial \vec{B} \over \partial t}
\label{eq:A.2}\\
\nabla\times\vec{B} &=& {\partial \vec{E} \over \partial t} +
e \dot{\vec{y}} (t) \, f(\vec{x}-\vec{y})
\label{eq:A.3}\\
\nabla\cdot\vec{E} &= e \, f(\vec{x}-\vec{y})
\label{eq:A.4}\\
{\rm and~~~~~~} \nabla\cdot\vec{B} &= 0 ~~.
\label{eq:A.5}
\end{eqnarray}
In the Coulomb gauge, the scalar potential is
\begin{equation}
\phi(\vec{x}) = {e \over 4\pi} \int {f (\svec{x}'-\vec{y}) \over
|\svec{x}'-\vec{x}|} d \svec{x}'
\label{eq:A.6}
\end{equation}
and the force coming from it is
\begin{equation}
\vec{f}_\phi = {-e^2 \over 4\pi} \int f(\vec{x} - \vec{y})
\nabla_x \int {d \svec{x}' f (\svec{x}'-\vec{y}) \over |\vec{x}-\svec{x}'|}
\, d\vec{x} = 0
\label{eq:A.7}
\end{equation}
identically.

The vector potential is transverse, and given by
\begin{equation}
\nabla^2 \vec{A} - {\partial^2 \vec{A} \over \partial t^2}
= - e \dot{\vec{y}}_\perp (t) \, f(\vec{x}-\vec{y}(t))
\label{eq:A.8}
\end{equation}
where $\dot{\vec{y}}_\perp$ is defined by Fourier transforming (\ref{eq:A.8}):
\begin{eqnarray}
- \vec{k}^2 \vec{a}(\vec{k}) 
- {\partial^2 \vec{a}(\vec{k}) \over \partial t^2}
&=& - e \dot{\vec{y}}_\perp \, f(\vec{k}) e^{-i \vec{k} \cdot \vec{y}}
\label{eq:A.9}\\
{\rm and~~~~~~} \dot{\vec{y}}_\perp &= \dot{\vec{y}} - \hat{k} \hat{k}
\cdot \dot{\vec{y}} ~~.
\label{eq:A.10}
\end{eqnarray}
Here,
\begin{eqnarray}
\vec{a}(\vec{k}) 
&=& \int e^{-i \vec{k} \cdot \vec{x}} \vec{A} (\vec{x}) d\vec{x} \nonumber \\
{\rm and~~~~~~} f(\vec{k}) &=& \int e^{-i \vec{k}\cdot \vec{x}} \,
f(\vec{x}) \, d\vec{x} ~~.
\label{eq:A.11}
\end{eqnarray}
We solve (\ref{eq:A.9}) for $\vec{a}$ with the boundary condition (for convenience only)
\begin{equation}
\vec{a}(\vec{k},t)=0 ~~,~~~~ t = 0 {\rm ~~~and~~~}
\dot{\vec{a}}(\vec{k},t)=0 ~~,~~~~ t=0 ~~.
\label{eq:A.12}
\end{equation}
The solution is
\begin{equation}
\vec{a} (\vec{k},t) = e \int_0^t dt' \,
{\sin k (t-t') \over k} \, \dot{\vec{y}}_\perp (t')
\, e^{-i \vec{k}\cdot\vec{y}(t')} f(\vec{k})
\label{eq:A.13}
\end{equation}
so with
\begin{eqnarray}
\vec{A}(\vec{x},t) &=& {1\over (2\pi)^3} \int d\vec{k} \, 
e^{i\vec{k}\cdot\vec{x}}
\vec{a}(\vec{k},t), 
\label{eq:A.14}\\
m_0 \ddot{\vec{y}} &=& e \vec{E}^e(t)
+ e^2 \int_0^t dt' \int {d \vec{k} \over (2\pi)^3} \, 
e^{i \vec{k} (\vec{y}(t)-\vec{y}(t'))} |f(\vec{k})|^2
\label{eq:A.15}\\
&&\hbox{\qquad}
\left\{ -\cos k(t-t') \dot{\vec{y}}_\perp (t')
+ \dot{\vec{y}}(t) \times 
\left( i \vec{k} \times \dot{\vec{y}}(t') \right)
{\sin k(t-t') \over k} \right\} ~~.
\nonumber
\end{eqnarray}
This is an exact integral equation for the electron motion.   If we
replace $m_0 \ddot{\vec{y}}$ by
$m_0 {d \over dt} {\dot{\vec{y}} \over \sqrt{1-\dot{\vec{y}}^2}}$
it would be a relativistic equation were it not for the cut-off function
$|f(\vec{k})|^2$.  If we set 
$|f(\vec{k})|^2 = 1$,
the equation becomes divergent (near $t'=t$)
and has no solution.   In view of these difficulties it makes no sense
to study this equation further, or to try to make it relativistic.
However, in the weak field limit we can and will solve the equation exactly, 
in order to study the way the run-away solutions appear.  In this limit
$y(t) - \vec{y}(t')$ 
and $\dot{\vec{y}}$
are of first order in $\vec{E}^e$ and ${\dot{y}}^k \times {\dot{y}}^\ell$ is of second
order.   Therefore the equation becomes
\begin{equation}
m_0 \ddot{\vec{y}} = 
e \vec{E}^e
- e^2
\int_0^t dt' \, 
\int {d \vec{k} \over (2\pi)^3} \,
\cos k(t-t') \, |f(\vec{k})|^2 \,
\dot{\vec{y}}_\perp (t')
\label{eq:A.16}
\end{equation}
and can be solved by Laplace transform.
Remembering our boundary conditions
$\vec{y}(0) = \dot{\vec{y}}(0)=0$,
we have, for $\vec{y}(\omega)$,
\begin{eqnarray}
-m_0 \omega^2 \vec{y}(\omega) 
&=& e \vec{E}^e(\omega)
+ {2\over3} e^2 \int {d \vec{k} \over (2\pi)^3} 
| f(\vec{k}) |^2
( -i \omega \vec{y}(\omega))
\int_0^\infty dt \, e^{i\omega t} (-\cos kt) \label{eq:A.17} \\
&=& e \vec{E}^e(\omega)
+ {2 \over 3}\omega^2 e^2 
\int {d \vec{k} \over (2\pi)^3}
{|f(\vec{k})|^2 \over k^2 - \omega^2} \vec{y} (\omega) ~~,
\label{eq:A.18} 
\end{eqnarray}
where
\begin{equation}
\vec{E}^e(\omega) = \int_0^\infty dt \, e^{i\omega t} \vec{E}^e(t)~~.
\end{equation}
So we find
\begin{equation}
\vec{y}(\omega) = 
{- e \vec{E}^e(\omega) / \omega^2 \over m_0 +  {2 \over 3} e^2 \int 
{d \vec{k} |f (\vec{k})|^2 \over (2\pi)^3 (k^2 - \omega^2)}}
{~~}.
\label{eq:A.19}
\end{equation}

$\vec{y}(\omega)$ in (\ref{eq:A.19}) has no upper half-plane singularities for $m_0 >
0$.  This follows from the observation that the imaginary part of
${1\over k^2 - \omega^2}$ is
\begin{equation}
{\rm Im~} {1\over k^2 - \omega^2} =
{2 \omega_1 \omega_2 \over (k^2 - (\omega_1^{\,2} - \omega_2^{\,2}))^2
+ 4 \omega_1^{\,2} \omega_2^{\,2}} ~~,
\label{eq:A.20}
\end{equation}
which is different from zero in the upper half $\omega$ plane unless $\omega_1 = 0$. 
Here
$\omega_1$ and
$\omega_2$ are the real and imaginary parts of $\omega$. 
When $\omega_1 = 0$, the denominator in (\ref{eq:A.19}) is always positive.  Thus
there are no run-aways.

Renormalization changes that.  Replace the integral
$$
 {2 \over 3}e^2 \int {d\vec{k} \over (2\pi)^3} {|f(\vec{k})|^2 \over k^2-\omega^2}
$$
by
\begin{equation}
 {2 \over 3}e^2 \int {d\vec{k} \over (2\pi)^3} |f(k)|^2
\left( {1\over k^2 - \omega^2} - {1\over k^2} \right)
+ \delta m
\label{eq:A.21}
\end{equation}
where
\begin{equation}
\delta m = {2 \over 3} e^2 \int {d \vec{k} \over (2\pi)^3} |f(k)|^2 ~~,
\label{eq:A.22}
\end{equation}
and note that the integral in (\ref{eq:A.21}) is now convergent in the limit 
$|f|^2 \to 1$.
In that limit, the denominator $D$ in (\ref{eq:A.19}) becomes 
\begin{eqnarray}
D &=& m_0 + \delta m + {2\over3} e^2 \int {d\vec{k} \over (2\pi)^3}
{\omega^2 \over (k^2-\omega^2) k^2}
\label{eq:A.23}\\
&=& m + {2\over3} {e^2 \over 4\pi} i \omega 
\label{eq:A.24}
\end{eqnarray}
and the run-away pole appears at
\begin{equation}
\omega = i m \cdot {6\pi \over e^2} ~~.
\label{eq:A.25}
\end{equation}

\def\theequation{B.\arabic{equation}}%
\section{Run-away in non relativistic quantum theory}


In this appendix we show that the classical formula (\ref{eq:A.19})
holds exactly, including arbitrary external field strength, in a
non-relativistic quantum theory in which we neglect recoil.
The neglect of recoil has an internal consistency, since recoil
can only be properly taken into account in a fully relativistic theory.
The point of this exercise is to show that quantum theory in itself does
not cure the run-away problem, since it almost identically reproduces
the classical effect.

We can solve this model exactly since the neglect of recoil makes the
Heisenberg equations of motion linear, and hence soluble, with the same
solution as the classical case.  Note that the neglect of recoil,
{\it i.e.} of the ${\vec{y}}$ dependence of $\vec{A}(\vec{y})$, 
is exactly the approximation we made following (\ref{eq:A.16}) to arrive at a
simply soluble classical problem there.

The Hamiltonian is
\begin{equation}
H = {(\vec{p} - e \vec{A})^2 \over 2 m_0} +
\sum_{k,\lambda} k {(q_{k,\lambda}^2 + \pi_{k,\lambda}^2) \over 2}
- e \vec{y} \cdot \vec{E}^e (t) 
\label{eq:B.1}
\end{equation}
where $q_{k,\lambda}$ and $\pi_{k,\lambda}$ are radiation oscillators:
\begin{equation}
[p_{k,\lambda}, q_{k',\lambda'}]
= {1\over i} \delta_{k k'} \delta_{\lambda \lambda'}
\label{eq:B.2}
\end{equation}
and 
\begin{equation}
\vec{A} = \sum_{k,\lambda}
{q_k \vec{\epsilon}_k \over \sqrt{k}} ~~.
\label{eq:B.3}
\end{equation}
A cut-off function can be introduced here, as in the classical
calculation of Appendix A.  We save space and time by simply inserting
it into the final answer, (\ref{eq:B.17}).

The Coulomb Hamiltonian
\begin{equation}
H_C = {e^2 \over 4\pi} \int 
{f(\vec{x}-\vec{y}) f(\svec{x}'-\vec{y}) d \vec{x} d \svec{x}'
\over |\vec{x} - \svec{x}'|}
\label{eq:B.4}
\end{equation}
is independent of $\vec{y}$ and hence
does not enter into the equations of motion.  These are
\begin{eqnarray}
\dot{\vec{p}} &=& - \nabla_{\!y} H = e \vec{E}^e(t) \label{eq:B.5}\\
\dot{\vec{y}} &=& \nabla_p H = {\vec{p} - e \vec{A} \over m}
\label{eq:B.6}\\
\dot{\vec{q}}_{k,\lambda} &= &{\partial H \over \partial \pi_{k,\lambda}}
= k \pi_{k,\lambda} 
\label{eq:B.7}\\
\noalign{\noindent{\rm and}}
\dot{\pi}_{k,\lambda} &=& - {\partial H \over \partial q_{k,\lambda}}
= -k q_{k,\lambda} + \dot{\vec{y}} \cdot 
\left( e \, {\partial \vec{A} \over \partial q_{k,\lambda}} \right) 
~~. \label{eq:B.8}
\end{eqnarray}
Although (\ref{eq:B.5})--(\ref{eq:B.8}) are operator equations, we will in what follows
only need mean values.  We therefore can set all operators to zero at
$t=0$, reproducing the classical boundary condition of Appendix A.
With that understanding, we carry out a Laplace transform.  As before, we
call
\begin{equation}
A(\omega) = \int_0^\infty dt \, e^{i\omega t} A(t)
\label{eq:B.9}
\end{equation}
for any variable $A(t)$.  The resulting equations are
\begin{eqnarray}
-i \omega \vec{p}(\omega) &=& e \vec{E}^e(\omega) \label{eq:B.10}\\
-i \omega \vec{y}(\omega) &=& {\vec{p}(\omega)-e\vec{A}(\omega)\over m}
\label{eq:B.11}\\
-i\omega q_{k,\lambda} &=& k \, \pi_{k,\lambda} \label{eq:B.12}\\
-i\omega \, \pi_{k,\lambda} &=& -k \, q_{k,\lambda} 
- i \omega \vec{y} \cdot {e \vec{e}_\lambda \over \sqrt{k}}
\label{eq:B.13}
\end{eqnarray}
with solution
\begin{eqnarray}
\pi_{k,\lambda} &=& - {i \omega \over k} \, q_{k,\lambda} \label{eq:B.14}\\
q_{k,\lambda} &=& - {e \sqrt{k} \vec{e}_{k,\lambda} \cdot \vec{y} \over
i \omega (1-k^2/\omega^2)} \label{eq:B.15}\\
{\rm and~~~~~~} \vec{y} &=& - {e \vec{E}^e \over \omega^2} \,
{1 \over m_0 + {2\over3} e^2 \sum {1\over k^2 - \omega^2}} 
\label{eq:B.16}
\end{eqnarray}
or, with the introduction of the cut-off function,
\begin{equation}
\vec{y}(\omega) = 
{- e \vec{E}^e(\omega) \over \omega^2 
\left[ m_0 + {2\over3} e^2 \int 
{d \vec{k} \over (2\pi)^3} \,
{|f(\vec{k})|^2 \over k^2 - \omega^2} \right]} ~~,
\label{eq:B.17}
\end{equation}
the same formula we found for the classical theory.  Of course
$\vec{y}(\omega)$ here represents the mean value of the operator.  Note finally that
adding a harmonic binding potential does not remove the run-away pole.

\def\theequation{C.\arabic{equation}}%
\section{Leading Radiative Correction}

The amplitude we must calculate is
\begin{equation}
M^{\mu\nu}(p,q)=i\int d^4y < p'|(j^\mu(0)j^\nu (y))_+e^{iq\cdot y}|p>
\label{eq:C.1}
\end{equation}
taken in the forward scattering limit, $p'=p$, and evaluated by continuing $q^0=w$ to
the upper half plane, as shown in Eq.~(\ref{eq:3.13}).

In lowest order
\begin{equation}
M^{\mu\nu}=M_0^{\mu\nu}=-e^2\bar{u}\gamma^\mu\frac{1}{i\gamma\cdot(p+q)+m}
\gamma^\nu u+\mbox{crossed term}
\label{eq:C.2}
\end{equation}
This formula is equivalent to Eq.~(\ref{eq:2.31}), which results from taking the
$q_i=0,(\mu,\nu)=(i,j)$ limit of (\ref{eq:C.2}).

The possibility of a Landau like pole will appear in lowest order by the coherent
positive addition to (\ref{eq:C.2}) of a term coming from the next order in $\alpha$, and
going like $\frac{\mbox{log}\; \omega^2}{\omega}$ or
$\frac{\mbox{log}\;\omega^2}{\omega^2}$ at large
$\omega$.

The possible asymptotic covariant amplitudes are limited by the conservation law $q_\nu
M^{\mu\nu}=0$.  Those that contribute to $M^{\mu\nu}$ in the next order are
\begin{equation}
T_1^{\mu\nu}=\bar{u}\frac{(\gamma^\mu i\gamma \cdot q \gamma^\nu-\gamma^\nu
i\gamma \cdot q\gamma^\nu)}{q^2} \, u
\label{eq:C.3}
\end{equation}
which goes like $1/\omega$ at large $\omega$;
\begin{equation}
T_2^{\mu\nu}=\frac{1}{Eq^2} \left( p^\mu p^\nu - (p^\mu q^\nu + p^\nu q^\mu) \frac{q
\cdot p}{q^2} + g^{\mu\nu} \frac{(p \cdot q)^2}{q^2} \right)
\label{eq:C.4}
\end{equation}
and 
\begin{equation}
T_3^{\mu\nu}=\frac{m^2}{Eq^2} \left( \frac{q^\mu q^\nu}{q^2} - g^{\mu\nu} \right) \ \ ,
\label{eq:C.5}
\end{equation}
both of which go like $1/\omega^2$ at large $\omega$.

One finds easily, to order $1/\omega ^2$,
\begin{equation}
M_0^{\mu\nu}=e^2(T_1^{\mu\nu} + 2 T_2^{\mu\nu})  \ \ .
\label{eq:C.6}
\end{equation}

The next order correction comes from the diagrams in Fig.~2 and Fig.~3.  It is, again
to order $1/\omega^2$, and keeping only terms with a factor log $\omega^2$ at large
$\omega$:
\begin{equation}
\delta M^{\mu\nu} = \frac{2\alpha}{3\pi} \; \mbox{log}\; q^2 M_0^{\mu\nu} -
\frac{4\alpha}{3\pi} e^2 \mbox{log}\; q^2 T_2^{\mu\nu} + \frac{2\alpha}{3\pi} e^2
\mbox{log}\; q^2 T_3^{\mu\nu} \ \ .
\label{eq:C.7}
\end{equation}
The first term in (4.7) comes from the Feynman diagrams in Fig.~3, which in a
different context signal the Landau ghost.

Finally, we point out two curiosities that emerge on inspecting Eq.~(\ref{eq:C.7}).  First,
since $T_1 \sim 1/\omega$, we see that the leading term in $1/\omega$ has no first
order logarithmic radiative correction, as implied earlier by Eq.~(\ref{eq:2.37}).  Second,
we see that the radiative corrections to the $1/\omega^2$ terms in the amplitude
$T_2^\nu$ coming from the diagram of Fig.~2 precisely cancel those coming from the
Landau diagram of Fig.~3.  Whether these are more than a numerical accident is not
known to the author.

\end{document}